# Recovering Solar Toroidal Field Dynamics From Sunspot Location Patterns

Aimee A. Norton and Peter A. Gilman

National Center for Atmospheric Research,
High Altitude Observatory, PO Box 3000, Boulder, CO, 80307-3000

norton@ucar.edu

gilman@ucar.edu

## ABSTRACT

We analyze both Kitt Peak magnetogram data and MDI continuum intensity sunspot data to search for the following solar toroidal band properties: width in latitude and the existence of a tipping instability (longitudinal $m=1$ mode) for any time during the solar cycle. In order to determine the extent to which we can recover the toroidal field dynamics, we forward model artificially generated sunspot distributions from subsurface toroidal fields that we have assigned certain properties. Sine curve fitting of Kitt Peak magnetogram data provided an upper limit of 15° to the tipping amplitude but could not adequately separate the tip from the width of the toroidal band. We then analyzed two sunspot distribution parameters using MDI and model data: the average latitudinal separation of sunspot pairs as a function of longitudinal separation, and the number of sunspot pairs creating a given angle with respect to the E-W direction. A toroidal band of 10° width with a constant tipping of 5° best fits MDI data early in the solar cycle when the sunspot band is at high latitudes (>18.5°). A toroidal band of 20° width with a tipping amplitude decreasing in time from 5 → 0° best fits MDI data late in the solar cycle when the sunspot band is at low latitudes (<18.5°). Model data generated by untipped toroidal bands cannot fit MDI high latitude data using chi squared goodness of fit criteria and can fit only one sunspot distribution parameter at low latitudes. Tipped toroidal bands satisfy chi squared criteria at both high and low latitudes for both sunspot distribution parameters. We conclude this is evidence to reject the null hypothesis - that toroidal bands in the solar tachocline do not experience a tipping instability - in favor of the hypothesis that the toroidal band experiences an $m=1$ tipping instability for a significant portion of the solar cycle. Our finding that the band widens from ≈10° early in the solar cycle to ≈20° late in the solar cycle may be explained in theory by magnetic drag spreading the toroidal band due to altered flow along the tipped field lines. Higher $m$ modes, most notably $m=2$ and $m=6$, are apparent in MDI data, but further analysis is needed to determine this property in detail.

*Subject headings*: Sun: magnetic fields --- Sun: activity

# 1. Introduction

We are interested in what can be learned about the dynamical behavior of the interior toroidal magnetic field from the statistical study of solar surface magnetism. Through analysis of sunspot data over significant time periods, we attempt to partially recover certain properties of the interior toroidal field. We ask the following questions: How broad, or wide, is the toroidal band in latitude? Can we observe any instabilities that the band experience? Of particular interest is a possible tipping of the toroidal band with respect to the equatorial plane. This tipping has been predicted by recent theory using a nonlinear model of a 2D MHD tachocline (Cally et al 2003; Dikpati et al. 2004).

Cally (2001) extended the studies of Gilman & Fox (1997), Dikpati & Gilman (1999), and Gilman & Dikpati (2000) to include non-linear behavior of the joint 2D instability of differential rotation and concentrated magnetic fields. He showed that broad toroidal fields open up into a `clamshell' pattern after nonlinear evolution, whereas Cally et al. (2003) showed that narrow toroidal bands just tip about their axis if ≈100 kilogauss (kG) whereas deformation occurs as well as tipping for fields of a few tens of kG or less. These studies suggest that a toroidal band may experience an instability causing it to tip with respect to the equatorial plane. The amplitude of the tip depends upon the latitude of the band, but is predicted to be ≈10° (5°) for bands of 10° (20°) width located around 30° (i.e. in the sunspot band), and less for lower latitudes (Cally et al. 2003).

The average latitude and temporal frequency of sunspot eruption has been studied extensively. Its periodic nature characterizes the ≈11 year sunspot cycle often represented by the butterfly diagram. What do we know about the toroidal field from studies of sunspots? The spread in latitude as mapped out by the butterfly wings can be used as a first order approximation to the width of the toroidal band. Estimates by eye show the width of butterfly wings experience growth until sunspot maxima when they span 20-30° in latitude. If the sunspot source is a ring of toroidal field whose field strength function in latitude is Gaussian, then the FWHM of the toroidal field strength is ≈10°. Also, the toroidal magnetic field probably has a strength between 10 and 100 kG. The field strength is most tightly constrained by a need to have flux emergence at low-latitude (Choudhuri & Gilman 1987; Choudhuri 1989; D'Silva & Choudhuri 1993; Fan et al. 1993) in accordance with observed sunspot emergence. Flux tubes with strengths much less than 30 kG would experience strong rotational influences and rise parallel to the rotation axis, causing sunspots to emerge at latitudes greater than 45°, which is not observed. The toroidal band is assumed to be located underneath the sunspot belt that migrates from ≈30° latitude early in the sunspot cycle to the equator late in the cycle.

Most studies to date have focused on the axisymmetric properties of the solar cycle. However, non-axisymmetric properties such as the longitudinal frequency of sunspot eruption is crucial in studying toroidal field dynamics because the instabilities predicted to exist on the toroidal field are non-axisymmetric in nature. Evidence of repeated flux emergence at specific longitudes, a phenomenon often referred to as active longitudes, may be observational evidence of a non-axisymmetric instability of progressively increasing order on the toroidal field (detoma et al. 2000). A toroidal field tipped with respect to the equator would also cause the latitudinal eruption of sunspots to have a longitudinal dependence, longitudinal wavenumber of $m=1$ or higher.

Of course, characteristics of the flux emergence process, such as sparseness of data and varying degrees of non-radial flux emergence, pose a challenge for the recovery of interior toroidal field structure. A threshold number of active regions is required in order to outline the basic geometry of the toroidal field for specified configurations. In this paper, we search for signatures of toroidal field dynamics and explore the limitations of our analysis techniques through forward modeling of sunspot emergence statistics. We fit observed distributions of sunspot locations with forward modeling data. We also explore the effects of data binning on the analysis. We use chi squared goodness of fit tests to determine if there is substantial evidence to reject the null hypothesis that the toroidal field is untipped.

## 2. Instruments and Data

The Kitt Peak data are observations made with the National Solar Observatory/Kitt Peak (NSO/KP) vacuum telescope. Each of the synoptic maps represents one Carrington rotation of the sun and has 360 x 180 pixel elements; see Fig 1. Longitude increases from left to right incrementally by 1°. There are 180 rows representing equal steps in sine latitude. A correction for the solar B0 angle is applied. On the assumption that that the fields are radial to the solar surface, the observed fields are divided by the cosine of η.
Noisy values near the poles are detected and replaced by a cubic spline fit to plausible values in the polar regions. Three different instruments have been used to produce the data. Rotations 1645-1853 used a 512 channel diode array magnetograph using the 8688 Å and 10830 Å spectrum lines. Rotations 1855-1862 used a CCD spectromagnetograph and the 5507 Å spectrum line for magnetograms. Rotations 1863 to the present use the spectromagnetograph and the 8688 Å line for magnetograms. Specific corrections to the data due to discontinuity in instrument use can be found at the NSO/KP data archive website.

MDI is a filtergraph space-based instrument observing full disk images in the 6768 Å Ni-I line. MDI synoptic Carrington Rotation (CR) intensity charts are generated from averages of 3 continuum intensity observations at each location that have been collected over the course of a 27+ day solar rotation. Intensity images are usually observed 2-4 times per day. After correction for limb darkening, and for flat field, orbital and instrument variations, re-mapped images are combined so that near central meridian data from 3 images are combined by taking the median value at each location. The synoptic charts are interpolated to disk-center resolution, resulting in a 3600 x 1800 pixel map. The axes are linear in Carrington longitude (0.1 degree intervals) and sine latitude. The continuum intensity values in the maps are computed using all five MDI filtergrams: $I_c=2F_0+I_{depth}/2+I_{ave}$ where $I_{ave}$ is the average of the other 4 filtergrams, $F_{1-4}$. The line depth quantity is the continuum minus the line center intensity: $I_{depth}=[\ 2((F_1-F_3)^2+(F_2-F_4)^2)\ ]^{1/2}$. Utilizing a summed quantity has the advantage of canceling systematic errors as a function of solar velocity. It is estimated that the continuum intensity measure is free from Doppler crosstalk at the 0.2% level (Scherrer et al. 1995). Processing of the MDI data is done by the Solar Oscillations Investigator team at Stanford University.

We limit our analysis to the location of sunspots instead of using information from a greater range of magnetic structures. This was done for several reasons. First and most importantly, sunspots are considered to be the most obvious manifestation of the toroidal field. Second, smaller magnetic field elements including flux from decaying active regions do not exhibit equatorward migration throughout the solar cycle and are therefore thought to be disconnected from the toroidal field. The inclusion of smaller magnetic elements in this analysis could complicate interpretation by introducing effects from surface flow dynamics instead of toroidal field dynamics. Third, sunspot records are available for longer durations making extension of this work possible to include data from the late 1800's to present.

**3. Search for Tipping Instability: Sine Curve Fitting to Kitt Peak Magnetograms**

The observation that sunspot patterns are often suggestive of a tipped toroidal field, such as the pattern seen in Fig 1, combined with recent results from theory predicting a tipping instability, motivated us to search the data systematically for a tipped toroidal field.

The surface pattern traced out by a purely tipped band, $m=1$, would appear as a sine curve within a given hemisphere when plotted in a latitude/longitude format. We use sine curves to fit the location of strong magnetic flux in 30 years of Kitt Peak Carrington Rotation (CR) magnetogram data. Weak flux was removed by use of a threshold value, set to 150 Mx/cm2, which removed most polar field flux as well as decaying active region flux. Central latitude, amplitude of tip, and phase in longitude were all free variables within the sine curve fitting routine. The grid allowed for fitting was 0-30 in central latitude with a 1° resolution, 0-20 in tipping with a 1° resolution and 0-360 in phase with a 2° resolution.

Fig Fig 2 shows the amplitude and central latitude resulting from the sine curve fitting of the Kitt Peak magnetogram data for CR 1645-1991 in the North and South hemisphere. The top panel obviously shows the sunspot band migration to the equator for the three different solar cycles, Cycles 21-23, captured during this time period. The bottom panel in Fig Fig 2 shows the resultant tipping amplitude to be $\approx 15°$ early in the solar cycles gradually diminishing to zero at solar minima.

The phase results were not as stable as expected for a true fit to a tipping instability. Periods of 8-15 Carrington rotations showed stable phases, but otherwise the phase values were noisy. One would expect the tipping instability to be somewhat stable in phase even if it had a slow phase speed on the order of 4-5°/CR. The unstable phase values is an indication that the sine curve fitting routine is not consistently finding a tip. The tipping amplitude values returned by the sine curve fitting routine may be more closely related to the latitudinal width of the toroidal field at a given time in the solar cycle than to a true tipping amplitude. We need an analysis method to distinguish between the width and the tip of the band.

## 4. Forward Model Description

We assume a band of toroidal field with the following properties: latitudinal width, W, peak field strength, field strength distribution as a function of latitude, and degree of tipping with respect to the equatorial plane (degree of tipping is the same as amplitude of $m=1$ mode). Fig Fig 3 shows a schematic of this configuration where the band of toroidal field is tipped with respect to the equatorial plane. The width of the toroidal band is defined as the latitudinal extent from which sunspots emerge, as shown in Fig Fig 4. The profile of the toroidal magnetic field peaks at ≈100 kG and decreases to 40 kG at the `edges' of the toroidal band from which sunspots are produced; see Fig Fig 4. This toroidal field is a banded profile, in the terminology of the instability modeling of Dikpati & Gilman (1999). The field strength of a given sunspot is linearly related to the toroidal field strength at the position from which it emerged and ranges from ≈1350-3500 Gauss.

The dynamics of the band include: the rate the band moves equatorward, when and at what amplitude tipping occurs, and the phase speed of the pattern migrating in longitude. The width of the toroidal band is not dynamic in any given run of the forward model, even though in reality it may vary during the course of a solar cycle. Random number generators determine the location (latitude and longitude) where a sunspot appears along the toroidal ring. The random number generation distribution is uniform in longitude and Gaussian in latitude. An input file provides the number of sunspots as a function of time.

For example, assume a toroidal band lies in the solar tachocline with a width of 10°. The model begins with the band located at 25° North of the equator and takes 143 CRs (≈ 11 years) to migrate to the equator. Throughout this time it has a constant tip of 5°, a phase speed of 4° longitude/CR. Model output is then the sunspot data (numbers, locations and field strengths) for each time step.

By assumption, half of the sunspot eruptions produced by the model produce a single sunspot while the other half of sunspot eruptions produce a sunspot pair. The sunspot pair is assumed to have an average separation and tilt angle of 35 Mm and 4.3° (Howard 1996). The active region tilt angle is implemented as a function of latitude, tilt angle = 2° + 0.2° x latitude, with spots at higher latitudes having a higher tilt angle as described by Joy's law. A uniformly random distribution of twice the tilt angle value is added as noise to the tilt angles in order to mimic the large distribution of angles found in solar data.

Flux emergence trajectories are another model parameter. We include an emergence distribution, ED, defined as the variations in trajectories of rising flux tubes as a function of field strength caused by Coriolis forces. Simulations show that weak flux has a more non-radial flux emergence trajectory than strong flux (Choudhuri & Gilman 1987). If a weak flux rope began its rise from the same location as a strong flux rope, its emergent latitude on the solar surface would be poleward of the stronger flux rope's. Based on numerical simulations (Fan et al. 1994) and observational studies (Norton & Gilman 2004), we limit the latitudinal distribution due solely to flux emergence to be 5°.

Higher longitudinal modes, up to *m*=6, were also simulated with our forward model. They were pure modes meaning we did not generate data from toroidal bands experiencing a combination of both *m*=1 and another mode. When generating data for higher m modes, all parameters of the model were treated as described above and only the number of sine waves traced out in latitude/longitude increased (i.e. *m*=2 mode traced 2 sine curves, *m*=3 mode traced 3 sine curves, etc).

## 5. How Many Sunspots are Needed to Recover Toroidal Dynamics?

A threshold number of sunspots, $N_{crit}$, is necessary to recover the interior dynamics from surface patterns. Considering only the *m*=1 instability and fitting a sine curve as a function of longitude to model generated data, $N_{crit}$ is determined when the best fit recovers the phase, amplitude of tipping and width of the toroidal band. This value is related to the spatial Nyquist frequency of sampling necessary to recover a pattern. However, since our sunspots are generated at random latitude and longitude locations, the optimal spacings of sunspots is hardly ever achieved and the $N_{crit}$ is higher than the Nyquist number. Also, including increasingly complex processes of sunspot emergence, such as emergence distribution effects and Joy's law, in the model adds noise into the system and increases $N_{crit}$. Note that the $N_{crit}$ values summarized in Table 1 are representative values since the random nature of the sunspot location determination produces slight variations from these values in a single forward model run.

Table 1 summarizes how changes in model parameters affect the $N_{crit}$ needed to recover the tipping amplitude, phase speed and central latitude of the tipped toroidal field to within a degree of its input model amplitude using the sine curve fitting technique described in Section 2. The first five rows of Table 1 illustrate that if the tipping amplitude is greater than the width of the toroidal band, recovery of the toroidal field pattern is more easily achieved with a small $N_{crit}$. If the tipping amplitude is small compared to the width of the toroidal band, it is more difficult to recover the pattern.

Including the effect of emergence distribution, i.e. weak magnetic fields rise non-radially through the convection zone, increases the $N_{crit}$ needed to recover the pattern. Including Joy's law in the forward model also increases $N_{crit}$. Joy's law is the single most influential parameter in pattern recovery. Joy's law was implemented such that follower spots were at higher latitudes than leading spots and every spot produced by the forward model was part of a spot pair. For the values shown in Table 1, no single sunspots were produced. The angle the spot pair made in relation to the E-W direction was randomly distributed about an angle *unrelated* to the tipping angle.

## 6. How Does Model Output Compare with MDI Data?

Sunspots were hand-selected (331 and 321 sunspots in N and S hemisphere respectively) from MDI synoptic continuum intensity images from CR 1910-2003. Their locations were recorded as a function of time. We used this data to address the search for a tipped toroidal field using a different approach than the sine curve fitting.

### 6.1. Latitude Difference as a Function of Longitudinal Separation

In the analysis of MDI data, we ask: How often are sunspots on opposite sides of the sun found further apart in latitude than those found near each other? If a tipping exists, sunspots 180° apart in longitude should, on average, show greater latitudinal separation than sunspots found closer together in longitude. Fig 5 plots the average unsigned separation in latitude of any two sunspots as a function of their separation in longitude. MDI data are analyzed separately for the North and South hemispheres and also for high and low latitudes for CR 1910-2003.

### 6.2. Sensitivity to Bin Size due to Higher m Longitudinal Instabilities

We were surprised by the sensitivity of the MDI data to bin size of longitudinal separation of sunspots as seen in Fig 5. MDI data results look very different as seen in the panels of Fig 5 with bin sizes of 90, 45 and 20°. Our first analysis utilized a bin size of 30° and showed a significant slope to the line (indicating the existence of a tipping instability) but whose first data point in the lowest bin of 0-30° longitude separation seemed aberrant due to a larger slope between it and the second data point of 30-60° longitude separation.

Altering the implementation of Joy's Law and increasing toroidal band width within reasonable limits did not successfully fit the slope between the first data points. After changing the bin size of our analysis to 10°, we found the slope was indeed higher for the first 0-30° of longitude separation. A possible physical reason for this was the presence of instability of the toroidal field to higher longitudinal wave number m. Specifically, the $m=6$ mode has a periodicity of 60° and can be Nyquist sampled by 30° bin sizes. We then realized that to sample the $m=1$ mode exclusively, we should use bin sizes of 90° and to sample an $m=2$ mode we should use bin sizes of 45°, etc. Our *a priori* assumption that only the longitudinal $m=1$ mode would be present was false.

For comparison purposes, we forward modeled sunspot location data based on toroidal bands experiencing pure modes of $m=1,2,3,4,5$ or 6 (not combinations of modes). Fig Fig 6 shows model results at left as seen for bin sizes of 90, 45 and 20° for the pure $m=1$, 2...6 modes. Plots at the right show MDI data for the same bin sizes. Obviously, $m=1$ is well sampled with 90° bin size. $m=1$, 2 and 3 modes are well sampled with a 45° bin size and all modes \le 6 are sampled with a 20° bin size. The modes $m=1,2$ and 6 are apparent in MDI data as seen in Figs Fig 5 & Fig 6.

### 6.3. Separation of Data into High & Low Latitudes

Separating data into high and low latitudes may increase chances of a tip identification because tipping amplitude is predicted to be greater at high latitudes (Cally et al. 2003). High and low latitude bins correspond to time periods of Carrington Rotations 1910-1957, when the average latitude of sunspots was >18.5°, and Carrington Rotations 1958-2003, when average latitude of sunspots was <18.5°. Model data was also separated into two time periods corresponding to when sunspots appeared at average latitudes of > or < 18.5°. This manner of binning separates the time series into two parts of equal length, although more sunspots are present in the later sequence.

Fig 5 shows results from the North and South hemispheres as well as from high and low latitudes. It is obvious that the hemispheres are similar whereas the high and low latitude data are offset significantly in average latitude difference. For the rest of the paper, North and South hemispheric data are averaged together but high and low latitude data are shown separately.

To compare observed data with forward modeled results, model data are plotted in the three panels of Fig Fig 7 with a constant tip of 0, 5 and 10°. A 0° tip means the model toroidal field is not experiencing a tipping instability. Toroidal bands with widths of 10, 15, 20 and 25° are plotted in each panel. Over-plotted in Fig 5 are the results from MDI high and low latitude data for the 90° bin size. In Fig Fig 7, MDI data was best fit by model data with a tip of 5°. High latitude data was best fit by the 10° band width whereas low latitude data was better fit by a wider band. Exhaustive goodness of fit tests were conducted varying all parameters and using 185 forward models. Table 2 and §7 summarize the goodness of fit test results.

### 6.4. Sunspot Pair Angles

Another way to view the data is to calculate the angle relative to the E-W plane defined by any two sunspots. This quantity is similar to the tilt angle of a bipolar action region except that we calculate the angle for any two sunspots present in a hemisphere at a given time. Fig 8 & Fig 9 are histograms plotting the number of sunspot pairs that display a given angle for both the MDI data at high and low latitudes as well as forward model data.

Estimated errors for the MDI histogram values were calculated by taking the MDI pixel size (2″) translated to degrees for disk center longitude and latitude (≈0.2°) and recalculating the observed sunspot pair angle for each sunspot pair ±0.2° times a random number for both the latitude and longitude positions of each sunspot. This was done 20 times and the standard deviation of the number of sunspot pairs in each bin was taken as the error. Error bars are over-plotted on MDI data shown in the right panel of Fig 8.

The sunspot pair angle was not as sensitive to binning as the average latitude difference as a function of longitudinal separation. Fig 8 displays MDI data in the right panel and the forward model data for longitudinal modes $m=0,1...5$ in the left panel. $m=6$ was not plotted in Fig 8 as it becomes obvious that the higher m modes all map out roughly the same curve. Modes with higher m are not well differentiated by analysis of the sunspot pair angles. The sunspot pair angles from MDI data were slightly sensitive to bin size. Originally, we had analyzed the number of pairs within bins of 0.5°. The curves appeared too noisy with this bin size so we increased the bin size to 1.5°.

In Fig Fig 9, MDI data is plotted as solid black line in addition to modeling results for toroidal bands whose latitudinal widths are 10, 15, 20 and 25°. The panels represent modeling results when the toroidal band is tipped with respect to the equatorial plane with angles of 0, 5 and 10°. The number of sunspot pairs for each curve is normalized such that the highest value is 1 in order to comparatively plot the data. MDI error bars are not over-plotted in Fig Fig 9 in order to simplify the plots and allow trends in model data to be apparent.

By eye, the best fits to MDI data appear to be a tip of 5° and width of 10° for high latitude data and a tip of 0° and width of 25° for low latitude data. However, a wider band with tip of 0° and a narrower band with the presence of higher order modes is indistinguishable using this sunspot pair angle parameter. For this reason, we feel it is important that both the sunspot parameters (pair angle and the average latitudinal difference of sunspots near or far away from each other) are fit by the forward model data. Exhaustive goodness of fit tests are summarized in Table 2 and §7 along with discussion.

**7. Goodness of Fit Test**

We use a chi squared ($\chi^2$) goodness of fit test to identify which forward modeled data best captures the behavior of MDI sunspot location parameters. We determine the best fit of forward model data to four curves: the average latitude difference as a function of longitude separation of sunspots for both high and low latitudes (see Fig Fig 7), and the histogram of number of sunspot pairs as a function of sunspot pair angle for both high and low latitudes (see Fig Fig 9).

$\chi^2$ values represent the averaged sum of the squared observed minus estimated (model) value divided by the estimated error (local error bars). Therefore, $\chi^2$ is a value reported in units of the local error bar. If the model is correctly capturing the physical system, random fluctuations in the data should roughly match the local error bars and the sum of $\chi^2$ over N data points should be ≈N. The average value of $\chi^2$ should then be ≈1. If $\chi^2 \gg 1$, it means the model is not closely mimicking the data and is probably wrong. On the other hand, if $\chi^2 \ll 1$ it might indicate that the calculated uncertainties are too large (Press et al. 1992).

For completeness, we determined a $\chi^2$ value for 185 forward models for each of the four curves. The models were as follows: 5 models with *m*=0 for toroidal bands whose widths were 5, 10, 15, 20 and 25°, 15 models with *m*=1 for band widths of 5, 10, 15, 20 and 25° and constant tipping amplitudes of 5, 10, and 15°, 15 models with those same widths and constant amplitudes for modes of *m*=2, 3, 4, 5, and 6. All forward models were then rerun (with the exception of *m*=0 modes) with instabilities whose amplitudes were not constant but linearly decreased from maximum amplitudes when the band was at highest latitudes to zero at the last time step when the toroidal band reached the equator. The amplitudes were still 5, 10, and 15° but these represented maximum amplitudes that decreased linearly in time to zero. These `variable' amplitude modes were tested because theory predicts tipping to be maximum at high latitudes decreasing as toroidal bands moved equatorward (Cally et al. 2003).

Results from the $\chi^2$ test are summarized in Table 2. Only values for $m=1$ models are shown in this Table because our emphasis is the identification of the $m=1$ mode. We will discuss higher m modes in the last paragraph of this section.

The best fit to the high latitude MDI data from average sunspot latitude difference as a function of separation in longitude using 90° bin size is a toroidal band whose W=10° with constant amplitude Tip=5° providing a $\chi^2$=0.2. The best fit to the low latitude data is W=15° with variable amplitude Tip=5\rightarrow0° providing a $\chi^2$=0.03. To test the null hypothesis - that toroidal bands in the solar tachocline do not experience a tipping instability as modeled with $m=0$ - let us compare the best fit $\chi^2$ to those for $m=0$ modes. The smallest $m=0$ $\chi^2$ value for both high and low latitude data is 4. We conclude that the $\chi^2$ test strongly favors a tipped toroidal field and the null hypothesis is rejected.

The best fit to the high latitude MDI data from the histogram of sunspot pair angles is either a band whose W=10° and a constant amplitude Tip=5° providing a$\chi^2$=0.8, or a W=15° and a variable amplitude Tip=5° providing a x2=0.7. The best fit to the low latitude data is either an untipped band whose W=25° providing a $\chi^2$=0.2 or a W=15° and a variable amplitude Tip=5° providing a $\chi^2$=0.3. We emphasize that this sunspot pair angle quantity is not as robust in distinguishing modes or distinguishing between contributions from tip versus width as the average sunspot latitude difference quantity. For example, when fitting low latitude data using sunspot pair angles, 17 different constant amplitude models and 33 variable amplitude models produced a $\chi^2$<1. When fitting low latitude data for average sunspot latitude difference, only one model produced a $\chi^2$<1. One simply cannot distinguish between higher modes or a wider band with no tip versus a narrower band with a small tip using this parameter.

We are therefore conservative and make our conclusions based on which models satisfy the $\chi^2$ criteria for both parameters. We emphasize that the $\chi^2$ results for both sunspot location parameters are in agreement. It is simply that the histograms of sunspot pair angles are somewhat ambiguous in that many models can fit the curves. Model data generated by untipped toroidal bands cannot fit MDI high latitude data using $\chi^2$ goodness of fit criteria and can fit only one sunspot distribution parameter at low latitudes. Tipped toroidal bands satisfy $\chi^2$ criteria at both high and low latitudes for both sunspot distribution parameters. We conclude this is evidence to reject the null hypothesis - that toroidal bands in the solar tachocline do not experience a tipping instability - in favor of the hypothesis that the toroidal band experiences an $m=1$ tipping instability for a significant portion of the solar cycle.

Because our models do not have the complexity to produce data based on combination of m modes and because $m=1$ is obviously present in MDI data, conclusions can not be reached regarding the specific amplitudes of higher m modes. However, it is important to mention that no combination of widths and amplitudes of the higher m mode model data satisfied the $\chi^2$ goodness of fit criteria for both sunspot distribution parameters. We calculated $\chi^2$ for all model data fits to MDI sunspot latitude difference using bin sizes of 45 and 20° (see Fig 6) and still found $m=1$ to dominate the signal and be the best fit.

Song & Wang (2005), using a fast Fourier transform method and studying the longitudinal spacing of active regions, find $m=5$ and $m=6$ present in Kitt Peak synoptic data. While this is consistent with our results, we did not use Fourier transforms to search for the tipping instability because we found the results were dominated by the spatial scale of active regions that may be independent of toroidal field dynamics.

**8. Summary**

Sine curve fitting to Kitt Peak magnetogram data showed results suggestive of a maximum tipping amplitude of 15° early in the solar cycle; see Fig Fig 2. However, phase results suggest that the fitting routine was not identifying patterns coherent over the timescales anticipated. We judge that the width of the toroidal band could not be separated from the tip using this analysis technique.

Sine curve fitting the surface patterns created with our forward model showed Joy's law to be the single most influential parameter in pattern recovery; see Table 1. Two systematic effects in our implementation of Joy's law contributed to this: 1) Joy's law was always implemented such that follower spots were at higher latitudes than leading spots and 2) the angle the spot pair made in relation to the E-W direction was *unrelated* to the tipping angle. Therefore, centroids of active region groups should be used in future data analysis of this type in order to minimize the effect that Joy's law has on the recovery of toroidal dynamics.

Assuming an emergence distribution exists in solar data of 5° and Joy's law can be approximated with the distribution described in §4, then the modeling results as summarized in Table 1 suggest one needs 12 active region groups to accurately recover a tip of 5° if the width of the toroidal field is 10°. This number of sunspot groups in one hemisphere is likely to exist only at certain times during solar maxima if at all. Time averages must be used during other phases of the solar cycle in order to achieve the threshold sunspot numbers critical to recovering toroidal field dynamics, and this averaging may compromise precise detection of a tip.

The comparison of MDI sunspot position data, averaged over ≈ 3.5 years in each hemisphere, to simulated data from our forward model strongly supports sunspots being generated from a tipped toroidal band, width 10° experiencing a tip of 5° in amplitude at high latitudes (early in the solar cycle). We find this band widens to 20° and the tip gradually decreases to 0° as the band moves equatorward; see Figs Fig 7 and Fig 9. $\chi^2$ goodness of fit criteria are satisfied by tipped toroidal bands for high and low latitude for both sunspot parameters of average latitude difference as a function of longitude separation and number of sunspot pairs with a given angle relative to the E-W direction. Untipped toroidal band model data could only satisfy the x2 criteria for low latitude data for the sunspot pair angle data. We conclude that this is strong evidence in favor of toroidal bands in the solar tachocline experiencing an $m=1$ instability with an amplitude of ≈ 5°. This is not opposed to the theory as Cally et al. (2003) predict an ≈10° tip for a 10° band located at a latitude of 30°. If this tip decreases in time as the band progresses equatorward, then our average value may well represent the true solar value for this time period.

Separating MDI and model data into two time periods when the toroidal band is at high latitudes (>18°) and low latitudes (<18°) shows the toroidal band to be wider by ≈10° later in the solar cycle; see Figs Fig 5, Fig 7, Fig 8 and Fig 9. The band appears to be ≈10° wide at high latitudes and 20° wide at low latitudes. Models which include diffusion in the form of kinetic and magnetic drag (Dikpati et al. 2004) show that the magnetic drag can prevent the the extreme 'clamshell' pattern from occuring but does not stop tipping from developing in banded profiles. An interesting result from these studies is that magnetic drag can cause a spreading of the band. Our results show a widening of the toroidal band in time. This could be interpreted as a spread in the band due to the influence of magnetic drag altering the flow pattern along the tipped field lines.

We should mention a few factors which have been found, through modeling effforts, to suppress or significantly alter the tipping instability. Shallow-water models show that the existence of a prograde jet acts to stabilize the toroidal band and inhibit the tipping instability (Dikpati et al. 2003). Since we have found tipping, we infer that if there are prograde jets inside the toroidal ring, they must be sufficiently small such that tipping still occurs (Dikpati & Gilman 2001). Supporting this are the results of a helioseismic search for jets reported by Christensen-Dalsgaard et al. (2005) who found only stationary jets at very high latitudes but no migratory jets in sunspot latitudes.

Higher m modes, notably $m=2$ and $m=6$ are apparent in the MDI data and are simultaneously present with the $m=1$ mode; see Fig Fig 6 and discussion of higher orders in §7. Our model does not at this time have the complexity to simulate sunspot location data with a combination of longitudinal modes. We anticipate adding this complexity to the model and being able to approximate the amplitude of the higher order modes in the future. The indication that higher wave numbers are present is of interest and may be due to the toroidal field being on the order of 20 kG or even less.

Future research also needs to address symmetry about the equator. The dynamics of the North and South hemispheres are important in determining whether symmetric or anti-symmetric dynamics dominate the toroidal band instabilities. The use of more extensive sunspot records, such as the Mt. Wilson or Greenwich Observatory sunspot data, could provide better constraints on toroidal band dynamics.

This research was suported by NASA grant NNH04AA49I. We are grateful to P. Cally and M. Dikpati for discussions related to this topic. We thank the referee, Dr. John Lawrence, for comments leading to an improved paper. NSO/Kitt Peak data used here are produced cooperatively by NSF/NOAO, NASA/GSFC, and NOAA/SEL.

# REFERENCES

Cally, P., 2001, Sol. Phys., 199, 231

Cally, P., Dikpati, M., & Gilman, P.A., 2003, ApJ, 582, 1190

Christensen-Dalsgaard, J., Corbard, T., Dikpati, M., Gilman, P., & Thompson, M.J., 2005, PASP Monograph Series, Proc. 22nd International NSO/Sac Peak Workshop, Oct. 2004, eds: K. Sankarasubramanian, M. Penn & A. Pevtsov, in press

Choudhuri, A.R., & Gilman, P.A., 1987, ApJ, 316, 788

Choudhuri, A.R., 1989, Solar Phys., 123, 217

de Toma, J., White, O.R., & Harvey, K.L., 2000, ApJ, 529, 1101

Dikpati, M., & Gilman, P.A., 1999, ApJ, 512, 417

Dikpati, M., & Gilman, P.A., 2001, ApJ, 559, 428

Dikpati, M., Gilman, P.A., & Rempel, M., 2003, ApJ, 596, 680

Dikpati, M., Cally, P.S., & Gilman, P.A., 2004, ApJ, 610, 597

D'Silva, S., & Choudhuri, A.R., 1993, Astron. Astrophys., 272, 621

Fan, Y., Fischer, G.H., & DeLuca, E.E., 1993, ApJ, 405, 390

Fan, Y., Fischer, G.H., & McClymont, A.N., 1994, ApJ, 436, 907

Gilman, P.A., 2000, Solar Phys., 192, 27

Gilman, P.A., & Fox, P., 1997, ApJ, 484, 439

Gurman, J.B. & House, L.L., 1981, Solar Phys., 71, 5

Howard, R.F., 1996, Solar Phys., 167, 95

Norton, A.A. & Gilman, P.A., 2004, ApJ, 603, 348

Press, W.H., Teukolsky, A.A., Vetterling, W.T., & Flannery, B.P., 1992, ``Numerical Recipes in C: the Art of Scientific Computing, 2nd Edition'', Cambridge University Press

Scherrer, P.H., Bogart, R.S., Bush, R.I., Hoeksema, J.T., Kosovichev, A.G., Schou, J., Rosenberg, W., Springer, L., Tarbell, T.D., Title, A., Wolfson, C.J., Zayer, I., and the MDI Engineering Team, 1995, Solar Phys., 162, 129

Song, W., & Wang, J., 2005, ApJ, 624, L137

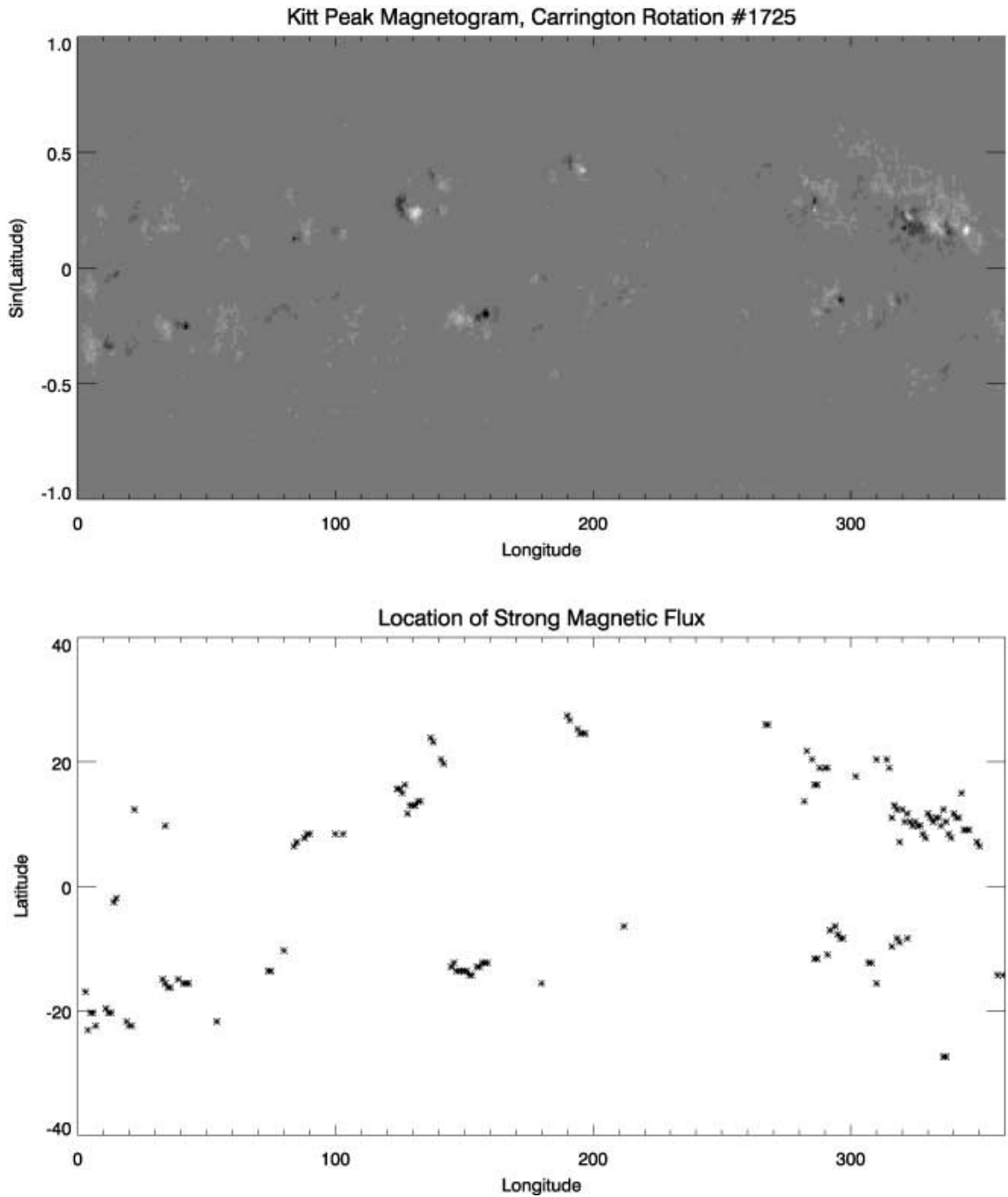

Fig. 1. - Kitt Peak data from Carrington Rotation 1725 is one example where the pattern traced out by surface magnetism is suggestive of a tipped toroidal field.

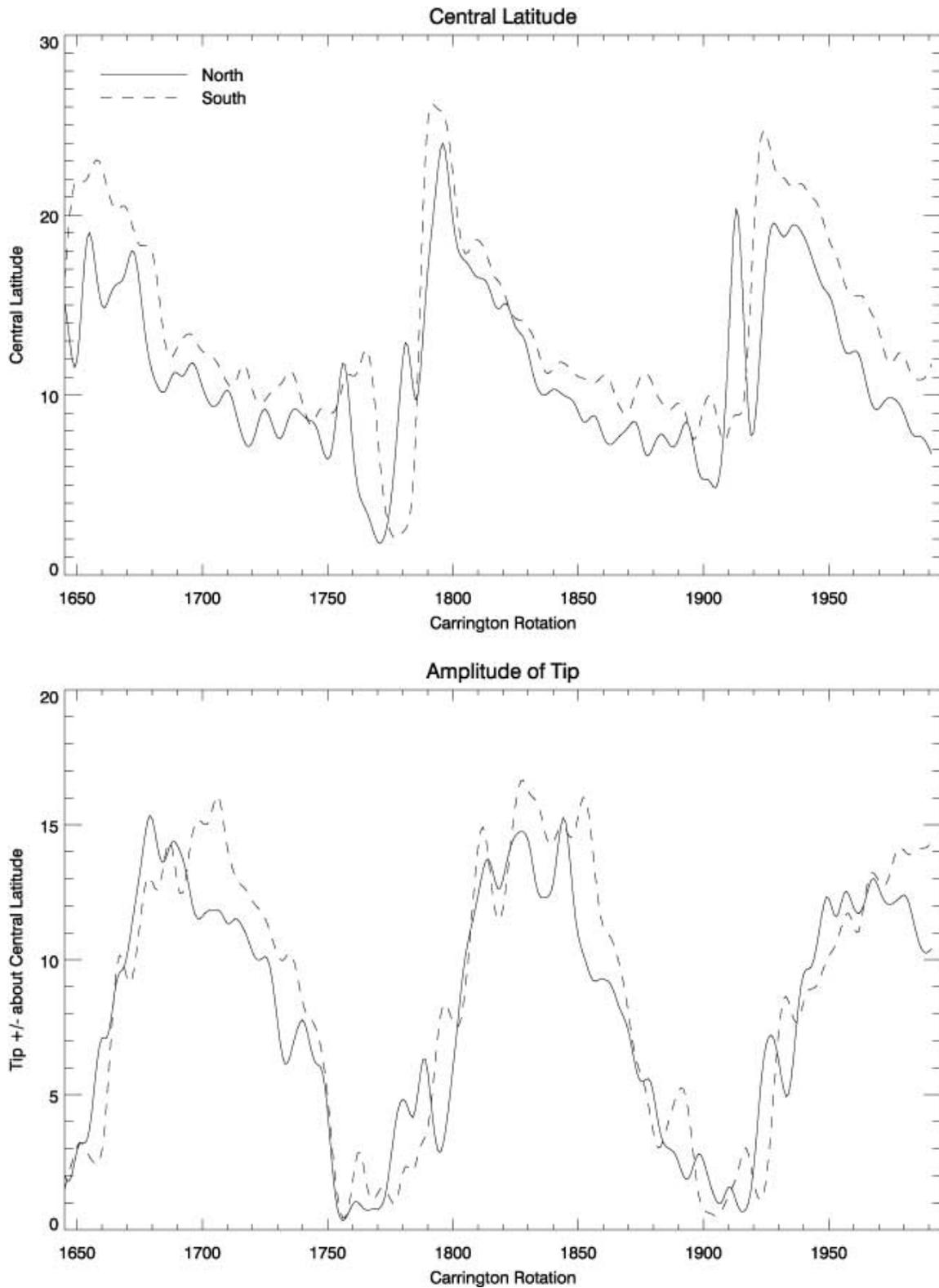

Fig. 2. - Central latitude (upper panel) and tipping amplitude (lower panel) resulting from sine curve fitting to Kitt Peak magnetogram data. Data was smoothed over 13 Carrington Rotations for plotting purposes.

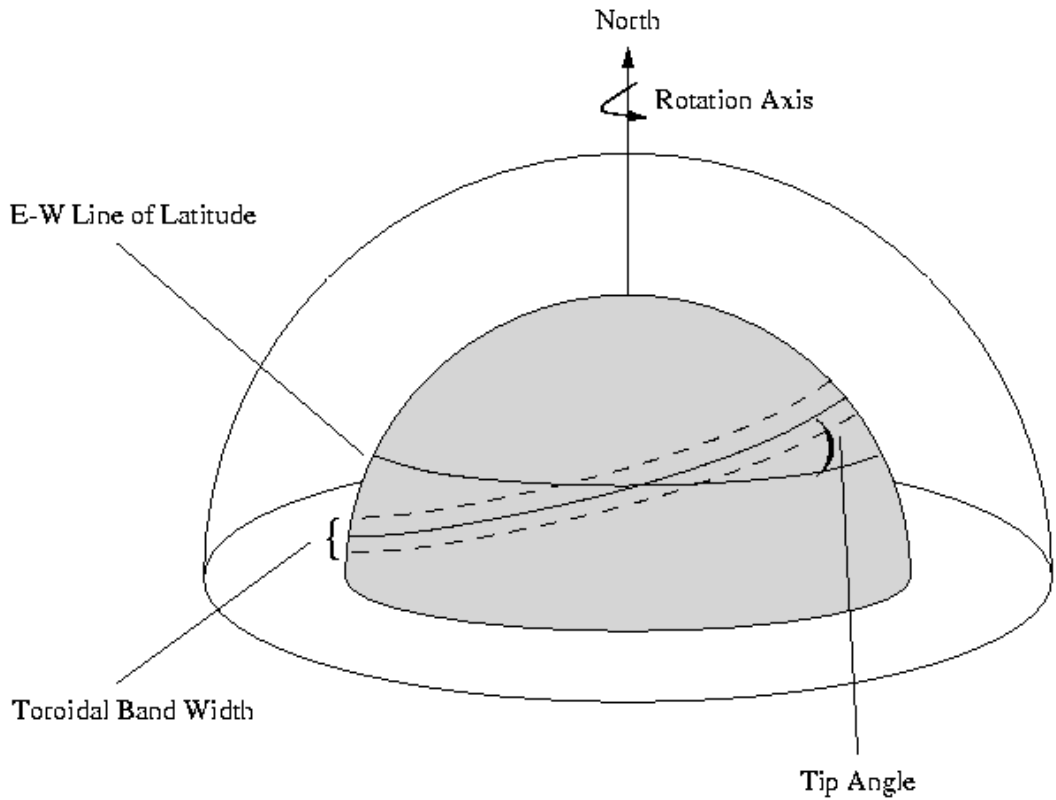

Fig. 3. - Schematic depicting a toroidal band of finite width tipped with respect to lines of constant latitude. Only the Northern hemisphere is drawn with the equator at the bottom of the image. The outer half-sphere represents the photosphere while the inner half-sphere (shaded) represents the base of the convection zone.

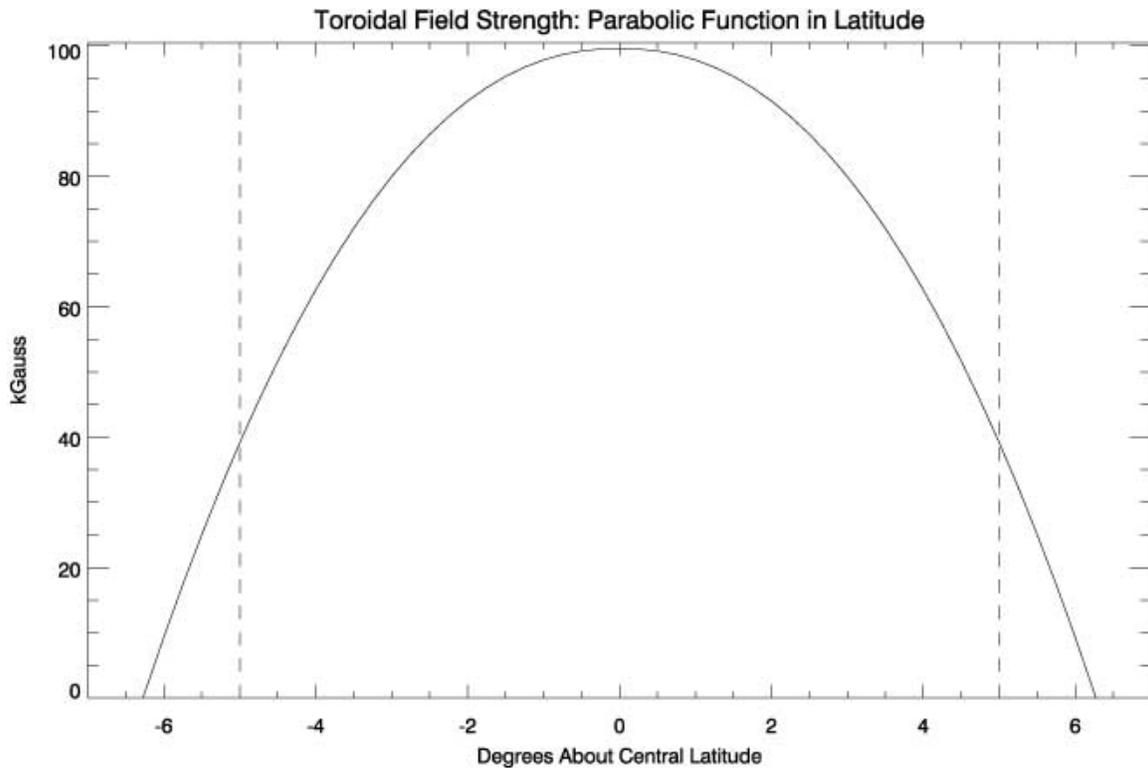

Fig. 4. - The profile of the toroidal field strength is a parabolic function of latitude. Sunspots emerge only between the dashed lines in order to limit the width, W, of the band and keep sunspot values on the order of observed values (1350 - 3500 Gauss). Width is defined as the latitudinal extent over which sunspots emerge. W=10° for this example.

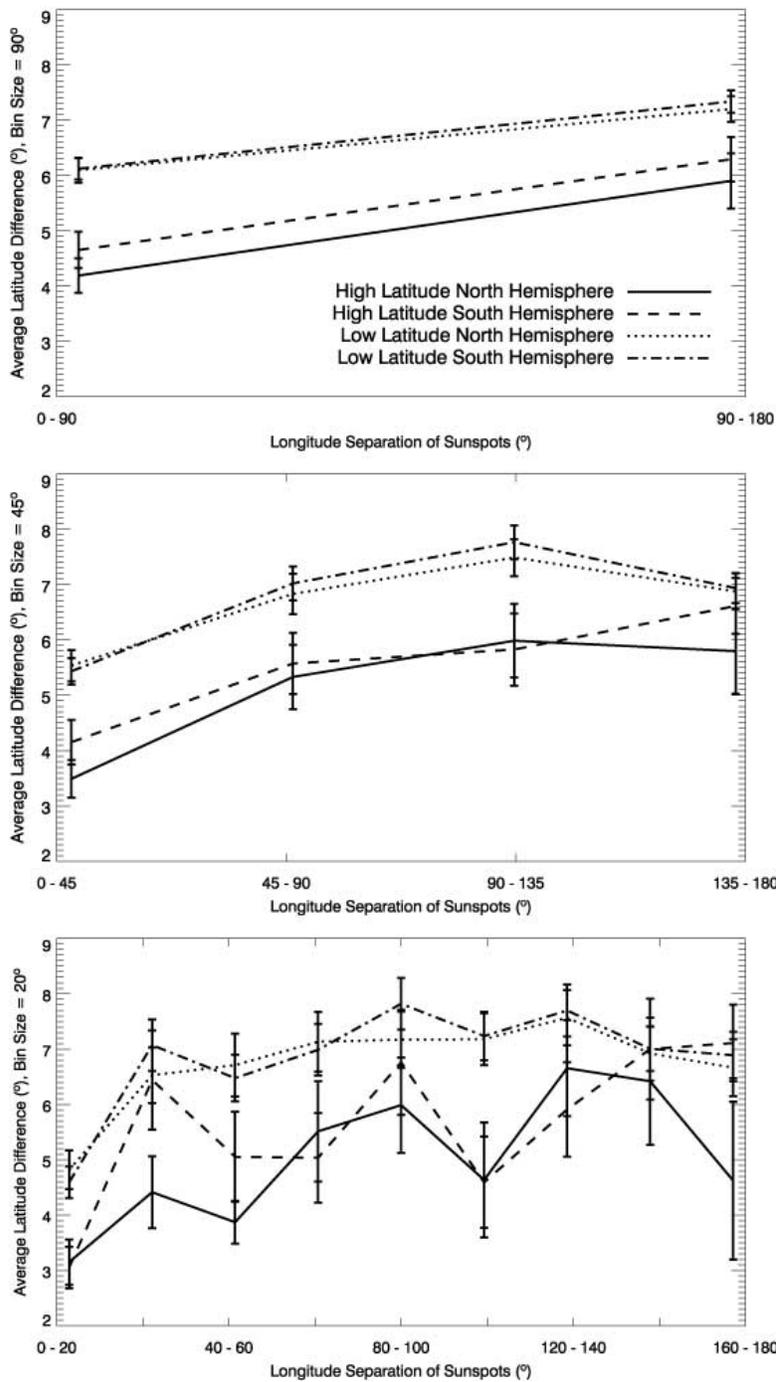

Fig. 5. - The average separation in latitude of any two sunspots is plotted as a function of their separation in longitude. MDI data is plotted for Northern and Southern hemispheric data both at high latitudes (early in solar cycle) and low latitude (late in solar cycle). Error bars are overplotted. The three panels from top to bottom illustrate the importance of bin size for sampling the longitudinal mode. Different bin sizes of longitude separation of 90, 45 and 20° are plotted in top, middle and bottom panels. The North and South hemisphere show strikingly similar trends. For the rest of the paper, North and South hemispheric data are averaged together but high and low latitude data are shown separately.

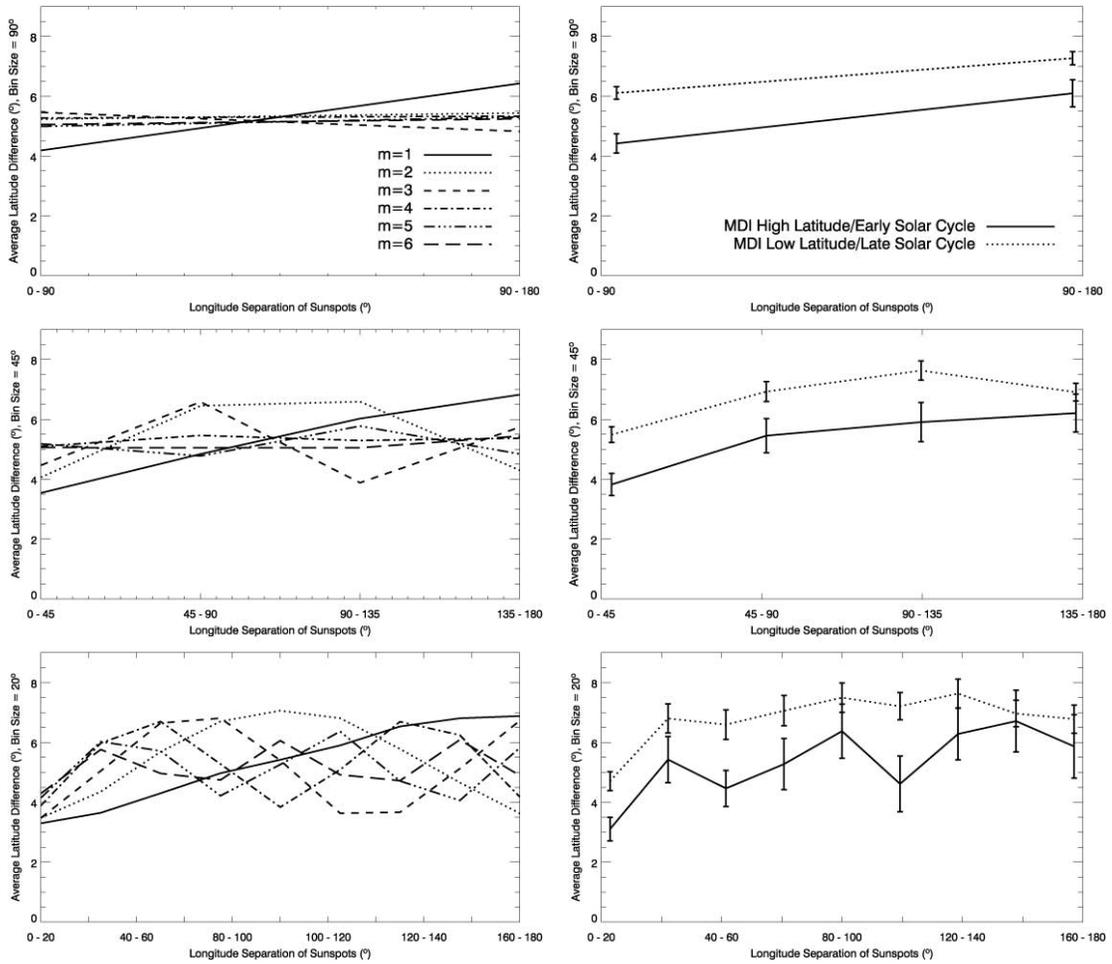

Fig. 6. - The average separation in latitude of any two sunspots is plotted as a function of their separation in longitude. MDI data is plotted as solid black line for high latitude data and dotted line for low latitude data in panels at right. Modeling results for toroidal bands whose latitudinal widths are 10, 15, 20 and 25° are plotted. The panels represent the modeling results when the toroidal band is tipped with respect to the equatorial plane with angles of 0, 5 and 10°.

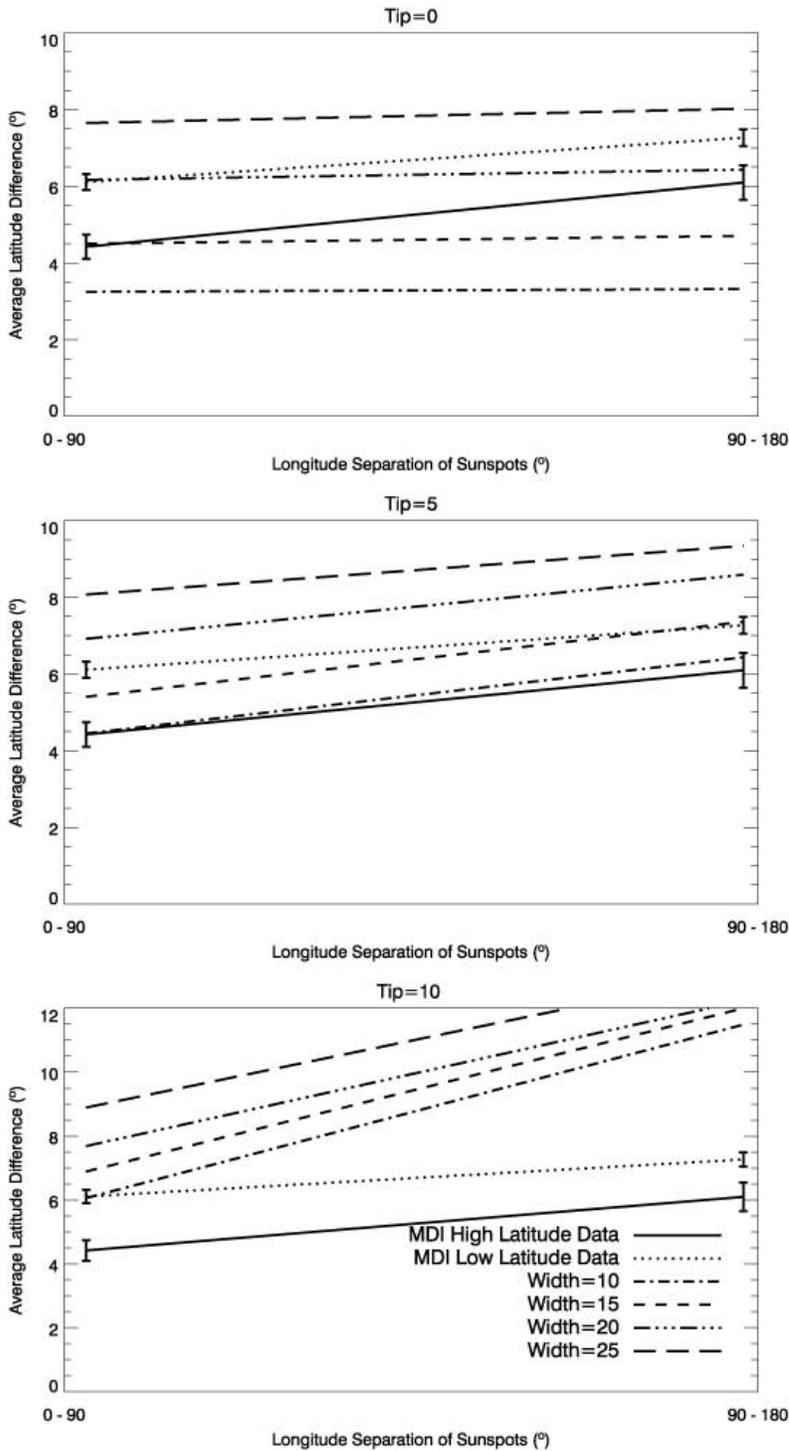

Fig. 7.- The average separation in latitude of any two sunspots is plotted as a function of their separation in longitude. MDI data is plotted as solid black line for high latitude data and dotted line for low latitude data. Modeling results for toroidal bands whose latitudinal widths are 10, 15, 20 and 25° are overplotted. The panels from top to bottom represent the modeling results when the toroidal band is tipped with respect to the equatorial plane with angles of 0, 5, and 10°. Results are shown with longitudinal bin size of 90° since that is optimum for sampling $m=1$ mode. }

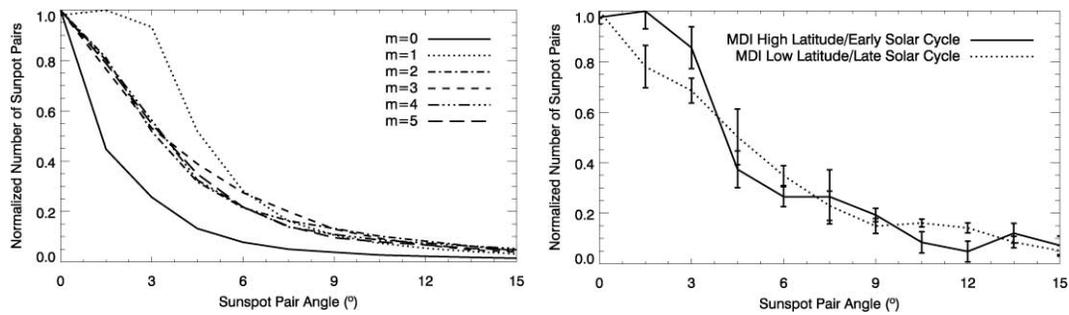

Fig. 8. - The number of sunspot pairs is plotted as a function of the angle that sunspot pair has with respect to the E-W direction. This is similar to the bipolar active region tilt angle except that it is not requires in our calculations that the two sunspots be within the same sunspot group or of opposite polarity. MDI data is plotted in the right panel as solid black line for high latitude data and dotted line for low latitude data. Modeling results for toroidal bands experiencing instabilities corresponding to longitudinal wave numbers m = 0,1...5 with a width of 10° and an amplitude of 5° is plotted in the left panel. Histogram bin size is 1.5°. Results were not sensitive to bin size.

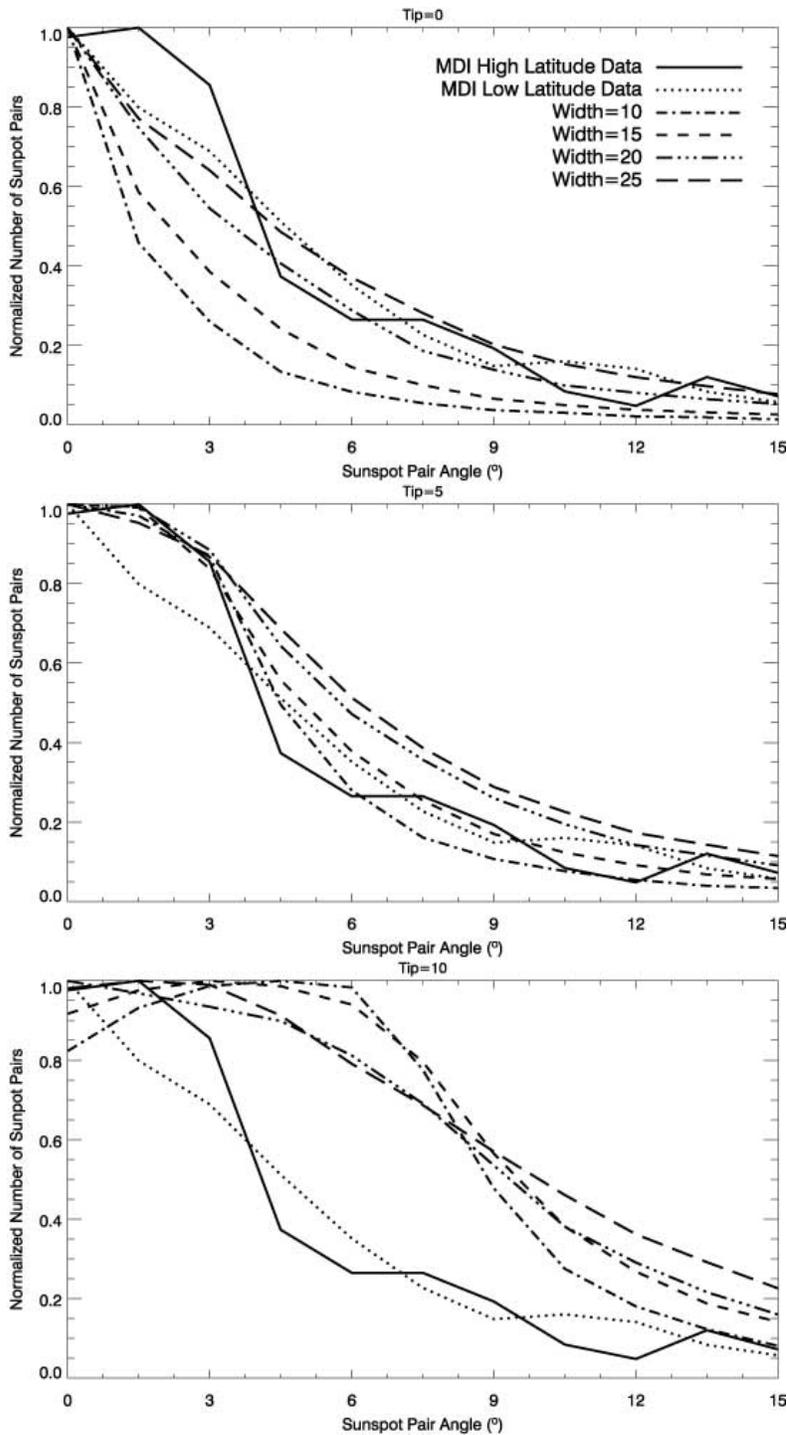

Fig. 9. - The number of sunspot pairs is plotted as a function of the angle that sunspot pair has with respect to the E-W direction. MDI data is plotted as solid line for high latitude data and dashed line for low latitude data. Modeling results for toroidal bands whose latitudinal widths are 10, 15, 20 and 25° are plotted. The panels represent the modeling results when the toroidal band is tipped ($m=1$ mode) with respect to the equatorial plane with tipping amplitude of 0, 5 and 10°.

Table 1. Number of sunspots necessary to recover interior toroidal dynamics.

| Width, W(°) | Tip, (°) | Emergence, ED(°) | Joy's Law | $N_{crit}$ |
|---|---|---|---|---|
| 5 | 10 | 0 | No | 3 |
| 10 | 5 | 0 | No | 5 |
| 10 | 10 | 0 | No | 4 |
| 10 | 15 | 0 | No | 3 |
| 15 | 5 | 0 | No | 7 |
| 20 | 5 | 0 | No | 9 |
| Including Emergence Distribution | | | | |
| 10 | 5 | 5 | No | 8 |
| 10 | 10 | 5 | No | 5 |
| 10 | 15 | 5 | No | 5 |
| 15 | 5 | 5 | No | 9 |
| 20 | 5 | 5 | No | 12 |
| Including Joy's Law, No Emergence Distribution | | | | |
| 5 | 10 | 0 | Yes | 16 (8 AR groups) |
| 10 | 5 | 0 | Yes | 18 (9) |
| 10 | 10 | 0 | Yes | 22 (11) |
| 15 | 5 | 0 | Yes | 32 (16) |
| 20 | 5 | 0 | Yes | 36 (18) |
| Including Joy's Law & Emergence Distribution | | | | |
| 5 | 10 | 5 | Yes | 22 (11) |
| 10 | 5 | 5 | Yes | 24 (12) |
| 10 | 10 | 5 | Yes | 28 (14) |
| 15 | 5 | 5 | Yes | 40 (20) |
| 20 | 5 | 5 | Yes | 42 (21) |

Table 2.  $x^2$ Goodness of Fit Test Results

| Tip(°) | W=5° | W=10° | W=15° | W=20° | W=25° |
|---|---|---|---|---|---|
| Sunspot Latitude Difference - High Latitude | | | | | |
| 0 | 53 | 22 | 4 | 9 | 39 |
| 5 | 1 | 0.2 | 5 | 24 | 62 |
| 10 | 63 | 76 | 92 | 115 | 165 |
| 15 | 313 | 307 | 315 | 370 | 384 |
| 5→0* | 8 | 2 | 1 | 17 | 49 |
| Sunspot Latitude Difference - Low Latitude | | | | | |
| 0 | 203 | 116 | 34 | 4 | 14 |
| 5 | 35 | 20 | 4 | 8 | 33 |
| 10 | 101 | 95 | 117 | 119 | 174 |
| 15 | 423 | 425 | 386 | 386 | 365 |
| 5→0* | 110 | 59 | 16 | 0.03 | 19 |
| Histogram of Sunspot Pair Angles - High Latitude | | | | | |
| 0 | 17 | 12 | 6 | 3 | 2 |
| 5 | 2 | 0.8 | 1 | 2 | 4 |
| 10 | 26 | 25 | 26 | 23 | 20 |
| 15 | 66 | 75 | 74 | 70 | 65 |
| 5→0* | 3 | 2 | 0.7 | 1 | 3 |
| Histogram of Sunspot Pair Angles - Low Latitude | | | | | |
| 0 | 15 | 10 | 4 | 1.3 | 0.2 |
| 5 | 4 | 2 | 0.8 | 2.3 | 3 |
| 10 | 22 | 23 | 26 | 22 | 21 |
| 15 | 60 | 68 | 61 | 55 | 42 |
| 5→0* | 7 | 4 | 1 | 0.3 | 1 |

*A tipping amplitude that varies from 5° at high latitudes decreasing linearly as a function of time to 0° at the equator.